\documentclass[journal,twocolumn]{IEEEtran}

\ifCLASSINFOpdf
\else
\fi
\let\labelindent\relax

\usepackage{graphicx}
\usepackage{adjustbox}
\usepackage[utf8]{inputenc}
\usepackage{booktabs}
\usepackage{subfigure}
\usepackage{tcolorbox}
\usepackage{amsmath}
\usepackage{amssymb}
\usepackage[numbers]{natbib}
\usepackage{paralist}
\usepackage{multirow}
\usepackage{xspace}
\usepackage{ulem}
\usepackage{color}
\usepackage{xcolor}
\usepackage[graphicx]{realboxes}
\usepackage{ifthen}
\usepackage{url}
\usepackage{fancybox}
\usepackage{enumitem}
\usepackage{listings}
\usepackage{balance}
\usepackage{ifthen}
\usepackage{graphicx}
\usepackage{amsmath}
\usepackage[linesnumbered,boxruled]{algorithm2e}
\usepackage{algorithm2e}
\usepackage{tablefootnote}
\usepackage{comment}
\usepackage{float}
\usepackage{algorithmic}
\usepackage{dirtytalk}
\usepackage{subfigure}
\usepackage{tikz}
\usepackage[colorinlistoftodos]{todonotes}

\usepackage{pgfplotstable}
\graphicspath{{pics/}}

\definecolor{codegreen}{rgb}{0,0.6,0}
\definecolor{codegray}{rgb}{0.5,0.5,0.5}
\definecolor{codepurple}{rgb}{0.58,0,0.82}
\definecolor{backcolour}{rgb}{0.95,0.95,0.92}

\lstdefinestyle{mystyle}{
	backgroundcolor=\color{backcolour},   
	commentstyle=\color{codegreen},
	keywordstyle=\color{magenta},
	numberstyle=\tiny\color{codegray},
	stringstyle=\color{codepurple},
	basicstyle=\footnotesize,
	breakatwhitespace=false,         
	breaklines=true,                 
	captionpos=b,                    
	keepspaces=true,                 
	numbers=left,                    
	numbersep=2pt,                  
	showspaces=false,                
	showstringspaces=false,
	showtabs=false,                  
	tabsize=2
}

\usepackage[english]{babel}

\DeclareGraphicsExtensions{.pdf,.jpeg,.png}
\begin{document}

\newtheorem{theorem}{Definition}[section]
	
	\newcommand{\eg}{e.g.,}
	\newcommand{\ie}{i.e.,}	
	\renewcommand{\lstlistingname}{Listing}

\newcommand{\boxedtext}[1]{\fbox{\scriptsize\bfseries\textsf{#1}}}
\newcommand{\nota}[2]{
	\boxedtext{#1}
		{\small$\blacktriangleright$\emph{\textsl{#2}}$\blacktriangleleft$}
}

\newcommand\autcomment[1]{{\textcolor{pink}{\textbf{}#1}}}

\newcommand\rev[3]{\textcolor{red}{\sout{#1}} {\textcolor{blue}{#2}} {}}

\newcommand\review[3]{\textcolor{red}{\sout{#1}} {\textcolor{blue}{#2}}{\todo{#3}}}


 \title{Challenges for Inclusion in Software Engineering: The Case of the  Emerging Papua New Guinean Society}

\author{
	\begin{tabular}{cccc}
		\multicolumn{3}{c}{Raula Gaikovina Kula, Christoph Treude, Hideaki Hata, Sebastian Baltes, Igor Steinmacher,} \\
		{Marco Aur\'{e}lio Gerosa, Winifred Kula Amini} \\
		{NAIST, University of Adelaide, Shinshu University, Federal University of Technology Paraná} \\ 
		{Northern Arizona University, PNG Digital ICT Cluster}\\
		{raula-k@is.naist.jp, \{christoph.treude, sebastian.baltes\}@adelaide.edu.au, hata@shinshu-u.ac.jp} \\
		{igor@utfpr.edu.br, Marco.Gerosa@nau.edu, winifred.amini@gmail.com}\\  
	\end{tabular} 
}
\maketitle

\begin{abstract}
Software plays a central role in modern societies, with its high economic value and potential for advancing societal change. 
In this paper, we characterise challenges and opportunities for a country progressing towards entering the global software industry, focusing on Papua New Guinea (PNG).
By hosting a Software Engineering workshop, we conducted a qualitative study by recording talks (n=3), employing a questionnaire (n=52), and administering an in-depth focus group session with local actors (n=5). 
Based on a thematic analysis, we identified challenges as barriers and opportunities for the PNG software engineering community. 
We also discuss the state of practices and how to make it inclusive for practitioners, researchers, and educators from both the local and global software engineering community.
\end{abstract}

\begin{IEEEkeywords}
Inclusion, emerging society, software engineering.
\end{IEEEkeywords}

\section{Introduction}
\label{sec:intro}
Advances in technology have the potential to transform societies~\cite{Castells2010NetworkSociety}.
As a society currently undergoing technological, economic, and social transformations, Papua New Guinea (PNG) is entering the digital age and showing signs of an emerging local software engineering community. In 2018, a local PNG software development team won the annual APEC App Challenge (\url{https://www.apec.org/Press/Features/2018/0530_app}).
Such progress paves the way for local economic development, with opportunities for young and talented individuals to develop software in areas such as natural resources, finance, healthcare, and education.
These developments also open the local society to the global software engineering (SE).

Following the work by Reijswoud et al.~\cite{van2009power} carried out in 2009, we set out to conduct a qualitative study with the local community in PNG, exploring the current situation and possible future directions in that emerging software development society. 
The goal of this study is to answer the research question, \textit{what are the challenges (as barriers and opportunities) for Software Engineering (SE) in an emerging society?}
To answer this question, we organised a workshop in PNG and conducted an empirical analysis based on three recorded talks, 52 questionnaire responses, and a focus group session with five participants.
Through thematic analysis, we derived five themes that affect software engineering in emerging societies such as PNG.
From that analysis, we distilled a set of local as well as global implications.
    
\begin{figure}
  \centering
  \includegraphics[width=.9\linewidth]{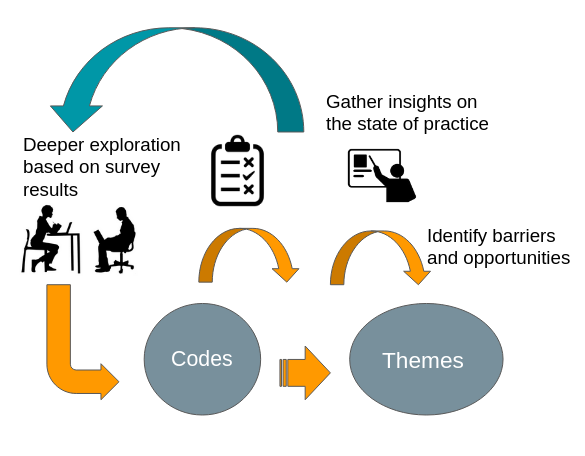}
  \caption{Overview of the thematic approach for the study}
  \label{fig:thematic}
\end{figure}

\begin{figure*}
  \centering
  \includegraphics[width=.8\linewidth]{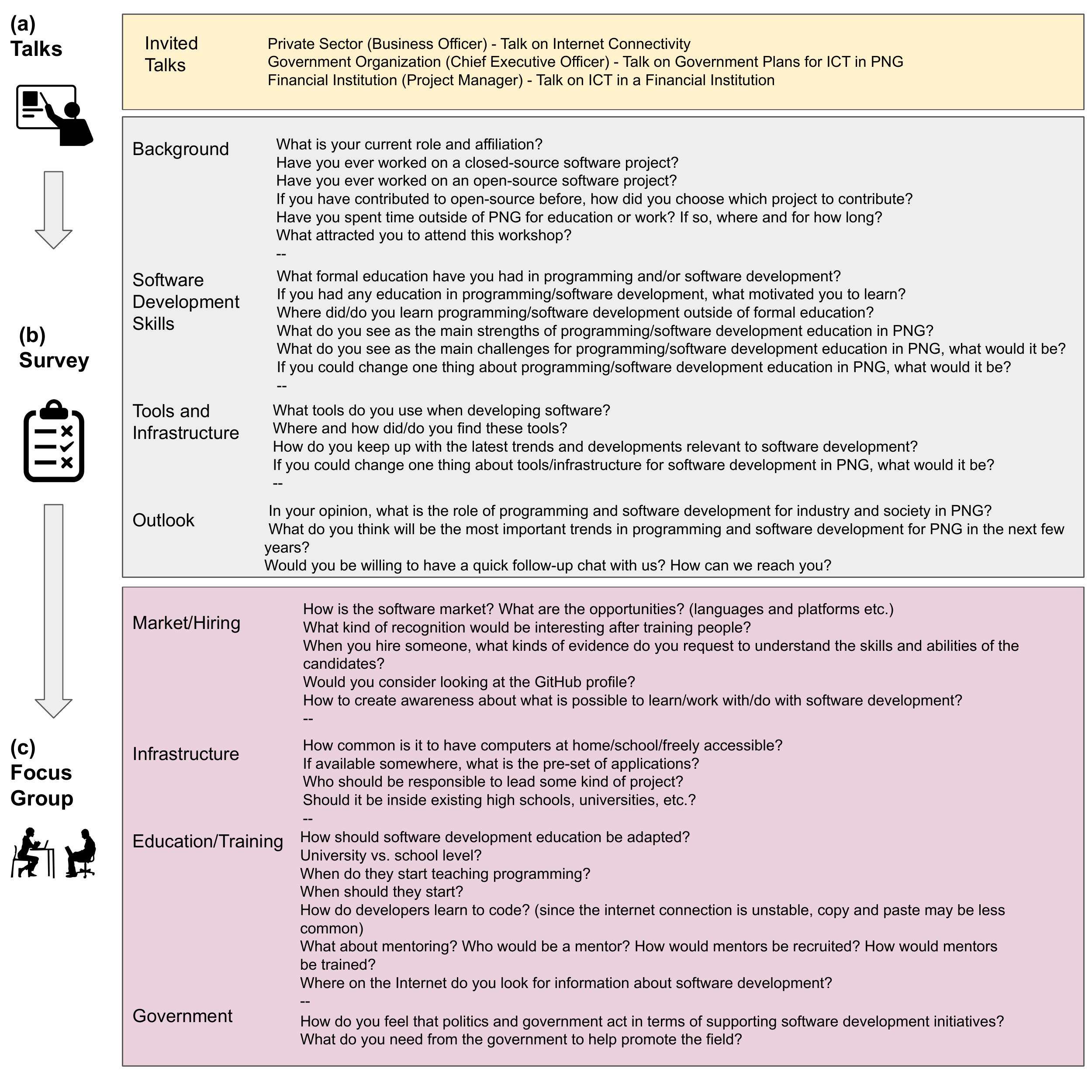}
  \caption{Summary of the three data sources, from which information was collected for the study.}
  \label{fig:datasource}
\end{figure*}

\section{Papua New Guinea as an Emerging Society}
\label{sec:back}

According to the World Bank \cite{web:PNGWorld}, PNG has a population of over 9.1 million (2020) with a diversity of geographic and natural resources.
The country also has an overall economic growth performance consistent with the real Gross Domestic Product (GDP) per capita, averaging 4\% since the mid-2000s.
Sound macroeconomic management and more efficient service delivery are critical to ensure that development benefits reach a higher number of Papua New Guineans, particularly given that 87\% of Papua New Guineans live in rural areas.
The growth trajectory and abundant resources provide a reliable platform for greater economic engagement with Asia and further abroad.
The country's economy is currently dominated by resource-driven sectors, such as agricultural, forestry, and fishing sectors and the minerals and energy extraction sector.
The World Bank report states that the population is highly dispersed and fragmented, as a result of the mountainous and archipelagic geography, low urbanization rate (13 percent), high ethnolinguistic diversity (840 distinct language groups), and social identities that are primarily small-scale, hence sharing the political economy characteristics of resource-rich states.

PNG’s emergence into the digital age may foster access to life-enhancing services in areas such as health and education. 
It may also catalyze innovation and economic growth, with the promise of new jobs and increased tax revenues (\url{https://www.gsma.com/r/mobileeconomy/pacific-islands/}).
 Like other emerging nations, PNG's Internet connectivity remains comparatively low, but is continuously improving.
From 2010 until 2013, the Internet penetration increased from 1.3\% to 6.5\%.
Since 2013, Internet penetration has jumped from 6.5\% to 30\% in 2018.
 This penetration is, however, significantly lower than in neighbour countries like Fiji (84\% in 2018). 
This situation is about to improve with a new submarine fibre-optic cable connection to Australia~(\url{https://oxfordbusinessgroup.com/news/how-png-achieving-faster-more-reliable-internet-access}).
The formal economy is divided into two sectors, state and non-state institutions.
However, PNG's informal economy was valued at 12 billion kina (US\$3.5 billion), which is equivalent to approximately one-fifth of the GDP and about 60\% of the non-resource GDP~\cite{web:informalsector}.
Considering the large number of people participating in the informal economy, the PNG government has begun to invest in Small to Medium Enterprises (SMEs) to advocate for financial inclusion~(\url{https://oxfordbusinessgroup.com/analysis/growth-unchained-new-technologies-coupled-infrastructure-upgrades-are-expanding-access-financial}).
 For example, since mid-2017, banks have actively promoted the distributed ledger technology blockchain for providing alternate forms of financial services.
Progress, however, has been slow.
As of mid-2018, only about 9.2\% of the population has access to some form of digital wallet.
 The PNG government has begun to invest in SMEs to advocate for financial inclusion.

\subsection{The Software Engineering Perspective: BRIDGES}
Based on the underlying research goal, we conducted a qualitative study involving local actors from the emerging local software engineering community.
To this end, we organized and hosted the International Workshop on BRIdging the Divides with Globally Engineered Software (BRIDGES) at \url{https://naist-se.github.io/BRIDGES2019/}. 
The workshop was run by the University of Papua New Guinea (UPNG) and Nara Institute of Science and Technology (NAIST). Recruitment was through UPNG, as well as open  invitations sent to the local high schools, and all state and non-state institutions located in Port Moresby.  We formally invited a representative from the PNG government. The workshop and survey were conducted in English as the main form of communication. The workshop was held for three days in Port Moresby, the capital city of PNG and the largest urban city in the Pacific region, with a population estimated at over 350 thousand.

\begin{table}
\caption{Demographics of invited speakers (S1-S3), questionnaire responses (Q1-Q52) and participants in the focus group (P1-P5)}
\label{tab:participants}
\begin{tabular}{@{}lcll@{}}
\toprule
 ID& Man/Women & Demographic & Sector \\ \midrule
 S1& Man & Business Officer & State Institution\\
 S2& Man & Chief Executive Officer & Government \\
 S3& Woman & Project Manager & Non-state Institution\\ \midrule
 Q1-Q23&-& High School Students & State Institution \\
 Q24-Q35&-& Professionals (IT)  & Non-state Institution \\
 Q36-Q45&-& Startup associates & SMEs\\
 Q46-Q52&-& Students (Undergrad) & State Institution \\\midrule
 P1& Woman & Managerial (IT) & Non-state Institution \\
 P2& Man & Managerial (Non-IT) & Non-state Institution\\
 P3& Woman & Student (Undergrad) & State Institution\\
 P4& Woman & Entrepreneur (CEO) & SMEs \\
 P5& Man & IT Consultant & Non-state Institution \\ \bottomrule
\end{tabular}%
\end{table}

\section{Study Design}
\label{sec:method}

Illustrated in Figure \ref{fig:thematic}, we use our research question to structure our analysis, based on the thematic analysis framework proposed by Braun and Clarke~\cite{braun2006using}.
Figure \ref{fig:datasource} shows the list of questions that were used in our study. Each data source is described below:
\begin{enumerate}
    \item \textit{Workshop Talks (n=3):}
    The BRIDGES Workshop talks given by local SE community members were recorded. 
    All three talks ranged from 25 to 30 minutes each and included interactive discussions with the audience afterwards.
    We recruited these three presenters as they represent different sectors in PNG, including government (S2) - Department of Higher Education Research, Science and Technology, non-state - an Internet Service Provider (S3), and financial institution (S1). 
    
    \item \textit{Questionnaire-based Survey (n=52):}
    We handed the questionnaire out to all BRIDGES workshop participants and to the participants of a startup meetup that was organized on the same day.
    The questionnaire was composed of workshop participants' demographics and their perceptions related to education, tools and infrastructure. 
    As shown in Table \ref{tab:participants}, we received 52 responses, with the majority of responses from high school or undergraduate students that attended the event. 
    The professionals include staff that work in the IT department, as well as managers and system analysts of their institutions. 
    \item \textit{Focus Group Discussion (n=5):} 
    After an analysis of the survey responses and workshop talks, the authors decided to organize a semi-structured focus group session. Participants were recruited from the workshop attendees and those that filled out the survey. The authors also made sure to incorporate diverse backgrounds and perspectives (professional and student, women and men, working in management and consulting, from IT and non-IT fields, and in large institutions and startups).
    This in-depth focus group ``round table'' discussion took place two days after we collected the survey responses, lasting 1 hour and 41 minutes.
\end{enumerate}

For data analysis, we used transcripts of the talks and the survey responses to gather insights on the state of practice, then used the focus group for a deeper exploration based on the survey results.
Our initial coding started with the analysis of the focus group, as it contained the largest content of information.
This resulted in a total of 79 codes, including codes such as ``not enough developers and designers'' and ``outside of Port Moresby, most education related to computers and IT is theory.''
The next round of coding then included the analysis and was merged with and reinforced by the codes generated from the workshop talks and the responses of the survey. 
The first four authors collaboratively conducted the initial round of analysis, while the remaining three authors were used as validity checks to ensure that the analysis was consistent throughout. 
 
 \begin{table}[btp]
\caption{Themes which emerged from the data analysis}
\resizebox{0.5\textwidth}{!}{%
\label{tab:themes}
\begin{tabular}{llrr@{}}
\toprule
&Theme & \# Focus Group  & \# Codes \\
& &  Contributions &  \\
\midrule
Barriers &Limited dedicated SE training & 4 & 25 \\
&Establish trust in local SE & 3 & 4 \\ \midrule
Opportunities&Investment potential in local SE  & 4 & 9 \\
&Market focused on & 4 & 29 \\
&customising off-the-shelf &  & \\
&Early stages of & 3 & 4 \\
&technological leapfrogging &  &  \\
\bottomrule
\end{tabular}
}
\end{table}

Table~\ref{tab:themes} summarises the themes which emerged from our analysis, together with the number of codes and number of focus group participants which contributed to each theme. 
As the table shows, each theme relates to codes that were assigned to quotes from at least three of the five focus group participants. 
An additional eight codes from the original annotation were used to inform the discussion of implications, cf.~Section~\ref{sec:implication}. 
Each theme was further augmented with data from the analysis of the survey responses and talks.

We sent a follow-up survey to 28 participants (i.e., the focus group participants, invited speakers, and survey participants who left their contact details) to further assess the credibility of our themes. 
Seven participants responded to the survey (response rate of 25\%).

\section{Findings}
\label{sec:find}
We identified the following themes as barriers and opportunities for the PNG SE community.

\subsection{\textbf{{Limited software engineering training (Barrier)}}}

With an adult literacy rate of 63.4\% as of 2015 (taken from \url{https://knoema.com/atlas/Papua-New-Guinea/topics/Education/Literacy/Adult-literacy-rate}), education unsurprisingly plays a large role in the advancement of PNG as an emerging society.
The needs around software development education were summarised by one of our focus group participants as follows:

\begin{quote}
\itshape
``I think we need specialised skills. From a graduate perspective, when we come out of uni, we only know the basics, we learn the general stuff---networking and programming, everything [...] From where I'm working now, there are IT people in the organisation, but I think they lack the specialised skill in software engineering so they can support the organisation and build something, build a system, instead of just the general knowledge of everything.'' --P3
\end{quote}



In the survey, we asked participants about one thing they would change regarding software development education in PNG.
Six professionals mentioned changing the education system either by teaching coding early (in primary or high school) or by offering higher-education courses dedicated to software development. 
Researchers and practitioners should be involved in designing an updated curriculum:

\begin{quote}
\itshape
``It's just they're learning stuff that's not up-to-date to where the industry is today. The other thing is, a research-backed and industry-backed curriculum that's developed so that when our students graduate, they go out with the skills that are relevant to the world today.'' --P4
\end{quote}

In an emerging society that has only recently reached the new frontiers of the digital age due to its developing education base, finding mentors within the country is challenging:

\begin{quote}
\itshape
``Nobody's done it before us. Without mentors, everything is pioneering, it's the first time.'' --P1
\end{quote}

Participants mentioned several online education websites, including FreeCodeCamp, O'Reilly, Udacity, and Coursera, which can serve as a substitute for formal education in some situations.




\subsection{\textbf{{Establish trust in local software engineers (Barrier)}}}
Taken together, the limited dedicated software engineering training and investment in the local software engineering workforce lead to trust issues in the abilities of the local workforce---a classic ``chicken-and-egg'' problem: Trust will only increase once the workforce is well-educated. Still, at the moment, investment goes to foreign entities rather than the local workforce, as P4 highlights:

\begin{quote}
\itshape
``If we tender for a project, like let's say the government, they're likely to get another firm outside, not a Papua New Guinean firm.'' --P4
\end{quote}



Developing trust in the local software engineering workforce is as much a cultural issue as it is an educational issue:

\begin{quote}
\itshape
``The problem we have not only in PNG but in the Pacific Islands is we don't trust our own [...] we need to pay for someone to be an international person to join us so that they can trust us. [...] Because we don't trust our own qualification. It comes down to that. [...] The mentality is the answers are always outside rather than inside, so they would pay larger amounts of money for someone outside rather than trusting someone inside.'' --P1
\end{quote}

\subsection{\textbf{{Market focused on customising off-the-shelf software (Opportunity)}}}
As a result of limited capacity as well as company and government policies to buy off-the-shelf software, the market for software developers in PNG is relatively small and provides more opportunities for customisation than developing software from scratch:

\begin{quote}
\itshape
``A lot of the bigger companies [...] buy off-the-shelf products. [...] The products are normally standard off-the-shelf, and then they would be customised, but usually the vendor would provide the software development to the company; and the in-house software development team would have very minimal software development. They would most likely develop the intranet for the internal company or the website.'' --P1
\end{quote}

Overseas companies, often from nearby Australia, are attempting to fill this gap, but this comes with costs related to software security, for example:

\begin{quote}
\itshape
``Most of the technology companies that do come in---most of the tenders for many of these large projects, not just large projects, but any sort of custom software development project---are overseas. There's a lack of local capacities, and support and maintenance are provided by those companies off-shore as well. [...] Security is a big issue. If people in the company don't understand how the system is built, especially if it's a custom one, then you never know how the system is properly protected.'' --P4
\end{quote}



While this could create a potential for the local software development market, such potential often remains untapped, and even startups focus on the customisation of existing technologies:

\begin{quote}
\itshape
``So the opportunity is for us to develop it based on our rules or the way we do business, but because we don't have enough software developers or designers, we can't be able to design systems that are fit for purpose.'' --P1
\end{quote}




\subsection{\textbf{Early stages of technological leapfrogging  (Opportunity)}}

Technological leapfrogging refers to the transit of countries from the condition of relative underdevelopment to that of an advanced industrial and technological state in a relatively short span of time~\cite{bhagavan2001technological}. 
This process typically follows three steps: (1) importing and absorbing highly modern technology, (2) replicating, producing, and improving the imported technology, and (3) moving on to innovations on one's own.
Our data suggest that PNG currently finds itself at the early stages of such technological leapfrogging.

\begin{quote}
\itshape
``Education is a big a thing. In PNG, amongst average Papua New Guineans, their idea of technology is just the Internet or Facebook or whatever they use.'' --P4
\end{quote}

To move further along the path to technological leapfrogging will require improved digital technology and digital literacy, as evidenced by this anecdote from one of our focus group participants:

\begin{quote}
\itshape
``Last year, when I was in another province of PNG, we ran a one week workshop for girls in ICT there. So we got girls from throughout that province from the high schools. And there were about 25 girls and three of the girls that attended had never touched a computer before. And they were literally scared to touch a computer because they thought they would break something. So not everyone has access to digital technology. [...] So access to digital technology is one, and then digital literacy is another thing.'' --P4
\end{quote}

A side effect of limited digital literacy is the tendency to try to run before one can crawl, e.g., by investing in the latest technologies despite a lack of foundational knowledge and experience. 
Investments into the latest technologies do not necessarily trickle down to the software engineering community:

\begin{quote}
\itshape
``A lot of our CEOs are traveling overseas, and they're reading a lot of these in-flight magazines. So there's a lot of content on Blockchain. So right now there's so much money being pumped in here to bring Blockchain software developers into the country. [...] But it's not being translated down, we don't understand how we could use Blockchain when the Internet's not that good, and we don't have the software engineering community.'' --P1
\end{quote}

A large barrier that is frequently mentioned in the talks, the survey responses, and the focus group is the high cost and low speed of the Internet infrastructure.
Mobile Internet access, which is most commonly used, costs around 150 Papua New Guinean Kina per month (approx. US\$45) for 30GB, unlimited access costs around 850 Kina per month (approx. US\$250).
Participants pointed to the general potential of ``open source software'' and ``learning online''.
\begin{quote}
    \textit{``There is no point in building a big highway (internet cable), [when] there are no cars (applications) to ride on that highway. We hope this workshop can give our people incentive to build software to put the cars on this road. We hope our people can build software that puts traffic on what we are building.'' --S1}
\end{quote}
S1 belongs to the state-owned entity that has been commissioned by the government to build a fibre-based network that will connect all 22 provinces across PNG, enabling the implementation of data centers that are expected to benefit the business, education, and healthcare sectors.

\subsection{\textbf{{Investment potential in local software engineers  (Opportunity)}}}
Boardrooms of Papua New Guinean companies were characterised as ``non-techy'' by the participants of our focus group:

\begin{quote}
\itshape
``If you see our executives, they are all fifty-plus, they are all old-school, they don't understand that tech plays an essential role in this day and age. [...] Our IT guys are lower down in the hierarchy, so they're not sitting with management.'' --P2
\end{quote}

As a consequence, companies do not invest in training through software engineering internships or joint courses with industry. 
While such internships exist in the mining sector, and mechanical and system engineering have internships and graduate programs, there are no internships in IT or software engineering: 

\begin{quote}
\itshape
``So when it comes to software, because the people at the top don't understand, we don't design the software according to what the requirements are, so it usually comes out not the way we expect it to.'' --P1
\end{quote}

\section{State of Practice}
\label{sec:implication}
As part of the follow-up survey, we find that participants strongly confirm statements relating to three of our five themes, but interestingly, there was no consistent agreement about the use of global technology, such as open source and global platforms like GitHub and Stack Overflow (early stages of technological leapfrogging) and whether dedicated SE training was up-to-date (training).
We now synthesize insights from our work to focus on lowering barriers and seeking opportunities for SE in PNG.

\subsection{For the PNG ICT Industry}
Our work covers the professionals that are both from the state, non-state and SMEs.
Similar to other resource-rich countries, non-state institutions rely on state-intuitions and donor organizations to drive change.
Hence, we identify three practices the all institutions could follow to enable inclusion into the global software engineering community.
This would be in the form of offering more (i) training, (ii) internships and (iii) lowering the barrier for SMEs.
Conducting short-term training to help onboarding local developers to create a local workforce capable of fulfilling the technical needs is an important step. It would directly address the theme of `limited dedicated SE training' and indirectly enable the market to not solely focus on customising off-the-shelf software if custom solutions can be built in PNG. Moreover, a qualified local workforce will help `establish trust in local SE' and lowering the barriers for startup companies will help with the `investment potential in local SE'.

At the level of PNG practitioners, our findings indicate that supporting local actors in an emerging society to become members of the global software development community can contribute to the sustainability of software engineering.
Such engagement could, for example, happen in the context of open-source software, through contribution and knowledge-sharing with global platforms such as GitHub and Stack Overflow.

\subsection{For the Global Research Community} 
Similarly to Craggs and Rashid \citep{Craggs:2019},
we identified trust as a crucial concept in our study. Further research is needed on how to establish trust in the local workforce in an emerging society, taking the important role of training investment into account.
More work is needed to understand what the predictors are for the success of such programs in a given society.
Inclusion in the global SE should be encouraged, for example, participation in international conferences leading to exposure to cutting edge technology.
Researchers should explore how global communication channels and social media can be leveraged as practices to help include the local SE communities from emerging societies. 
For example, PNG participants were invited to attend a Mining Software Repositories conference under the widening participation program (details at \url{https://2021.msrconf.org/attending/widening-participation-program}) for inclusion. 

\subsection{For Educators (both local and global)}
Our findings indicate a need to update the existing education curricula related to SE.
In-line with the PNG ICT industry implications, the focus for the state and non-state institutions should be on upgrading the capacity of local educators, and importing foreign education.
Non-state institutions should advice that education of the local ICT should be a higher priority than simply offshoring work, which becomes costly in the long run. 
Global initiatives such as Google Summer of Code (GSoC) are partly aimed at emerging societies, and have found large uptake in some of them.

\section{Outlook}
\label{sec:conclude}

Recent literature has recognized that existing software engineering (SE) processes, methods and practice are not suitable for the direct application in emerging countries~\cite{osman2012teaching}. Reports from Ethiopia~\cite{teka2016usability}, Iran~\cite{khaksar2015study}, Serbia~\cite{stefanovic2010ict}, and Sub-Saharan Africa~\cite{mursu2000information} highlight different problems that affect the success of software development industry in these countries; in particular they mention low infrastructure~\cite{teka2016usability, mursu2000information}, lack of financial resources~\cite{khaksar2015study, stefanovic2010ict}, and social and environmental factors~\cite{mursu2000information}. 
We can see that most of the reported problems are similar to those we uncovered in this paper. 

There are also initiatives in place aiming to strengthen the software and Information Technology industry in a way to benefit these countries. There are several ICT4D (Information and Communications Technologies for Development) initiatives aimed at bridging the digital divide by ensuring equitable access to up-to-date communications technologies. One example is the Health Information Systems Programme (HISP) for developing countries (\url{https://www.mn.uio.no/ifi/english/research/networks/hisp/}).

When we look at the literature focusing on software development industry in developing and underdeveloped countries, several studies show that government support is essential for the growth of a local ecosystem, by supporting education and infrastructure initiatives, for example~\cite{khaksar2015study, stefanovic2010ict}. Other authors specifically point to Open Source Software (OSS) as a trigger to the software industry in an emerging country~\cite{gakh2018open, camara2007information}. PNG has not embraced these advantages offered by OSS yet, as we could notice analyzing the work of Van Reijswoud et al.~\cite{van2009power}. In addition to the strategies mentioned above, we saw that one of the main strategies to foster the local software industry is reshaping education in emerging countries (e.g., \cite{osman2012teaching, garg2008software}.)

The emerging society of PNG has only recently reached the digital age frontiers, with early signs of a local SE community.
While we cannot claim generalisation of our results beyond PNG, we expect that our insights can be transferred to similar emerging societies. Our future work will focus on testing our assumptions in a wider context, especially in the area of Global Software Development, to solve trust issues in distributed teams, and communication issues across different cultures.

\section*{Acknowledgements}
We would like to thank all the participants and organizers of the BRIDGES at \url{https://naist-se.github.io/BRIDGES2019/}, from NAIST and the University of PNG. Furthermore, the event was supported by Oil Search, PNG DataCo, PNG Science and Technology Council Secretariat, Nasfund Papua New Guinea and PNG Digital ICT Cluster.

\bibliographystyle{IEEEtran}
\bibliography{icse-seis}

\end{document}